**DNNsolver: Accurate and Efficient Deep Learning Modeling of Polarization-Sensitive Diffractive Neural Networks**

*Shu-Yi Wang, Hong-Bo Sun*, and Linhan Lin**

S.-Y.Wang, H.-B.Sun, L.Lin
State Key Laboratory of Precision Measurement Technology and Instruments, Department of Precision Instrument, Tsinghua University, Beijing 100084, China

E-mail: hbsun@tsinghua.edu.cn, linlh2019@mail.tsinghua.edu.cn

## Abstract

Diffractive neural networks (DNNs), composed of cascaded diffractive layers, offer a promising platform for high-speed optical computing. However, accurate and efficient modeling of high-density, polarization-sensitive DNNs remains challenging. Conventional model based on thin element approximation (TEA) fails to capture electromagnetic coupling within layers, while full-wave simulations such as finite-difference time-domain (FDTD) methods are computationally expensive. Here, we propose DNNsolver, a deep learning-based electromagnetic surrogate model that learns the scattering response of diffractive layers. By predicting the scattering matrix rather than source-dependent output fields, DNNsolver decouples the incident wavefront from the model and could be applied to arbitrary input fields. A local scattering kernel and sliding-window strategy further allow efficient modeling of DNNs with arbitrary spatial sizes. For single diffractive layer under random speckle illumination, DNNsolver achieves an average mean squared error (MSE) of 0.0095 and about $10^6$-fold speedup compared with FDTD simulations. DNNsolver is further applied to the optimization of a trilayer polarization-multiplexed image classifier, which exhibits better agreement with FDTD simulations than its TEA-based counterpart. Our work provides an accurate and efficient framework for the design and optimization of high-density, polarization-sensitive DNNs.

## 1. Introduction

Diffractive nerual networks (DNNs) have emerged as an attractive optical computing framework that exploits optical diffraction and interference to implement neural network inference in physical domain [1]. A typical DNN is composed of cascaded diffractive layers that modulate the phase of transmitted wavefront. These diffractive layers are commonly modeled using the thin element approximation (TEA), which treats each layer as a two-dimentional phase modulator and greatly simplifies the forward propagation model. However, TEA is valid only when the system works under paraxial conditions and when the unit-cell period is much larger than the working wavelength, such as DNNs implemented with diffractive optical elements [2–5] or spatial light modulators [6–10]. When the unit-cell period approaches the operating wavelength without a corresponding reduction in the layer thickness, the accuracy of TEA model deteriorates [11–14]. Under such conditions, large-angle diffraction and strong electromagnetic coupling between adjacent unit cells become significant, especially for high-spatial-frequency fields generated within multilayer DNNs. This issue is further aggravated in DNNs based on low refractive index materials [15–17], which require thicker layers to provide sufficient phase modulation. Consequently, TEA fails to capture the underlying wave propagation physics, limiting the accurate design and optimization of high-density diffractive neural networks.

The finite-difference time-domain (FDTD) method, owing to its high accuracy in resolving full-wave electromagnetic interactions, has been widely employed in nanostructure simulation and optimization [18–21]. However, its prohibitive computational cost renders it impractical for training DNNs, where a single iteration commonly requires thousands of simulations. In addition, algorithms based on simplified version of Maxwell equations, e.g., beam propagation method (BPM) [22, 23] and wave propagation method (WPM) [24–26], enable much faster simulation with substantially reduced computational cost. Such efficiency, however, is achieved at the expense of modeling accuracy, particularly in scenarios involving strong diffraction or complex electromagnetic interactions. Such existing modeling approaches are fundamentally constrained by a trade-off between computational efficiency and physical accuracy.

Recently, deep learning driven models for electromagnetic simulation have emerged, offering substantially improved computational efficiency while maintaining high predictive accuracy. The mapping from structual geometries to output field distribution is learned using convolution neural networks [27–30] or Fourier neural operator based architectures [31]. However, these architectures inherently restrict the simulation domain to a fixed size, requiring new datasets and model retraining whenever the simulation region changes. Some other models tailored for metasurface overcome the size limitation with sliding window approach [32–35], while the local structural characteristics remain statistically consistent. Nevertheless, all the models mentioned above are generally developed for a single predefined incident field and cannot be directly applied to DNNs, where the incident wavefront at each layer is continuously reshaped by the preceding layers and could take arbitrary forms. This challenge fundamentally arises from the mismatch between the nonlinear representation capability of DNNs and the intrinsic linearity of optical system. Consequently, an efficient and accurate electromagnetic surrogate model capable of accommodating arbitrary incident wavefronts remains elusive so far, posing a critical obstacle to the scalable design and optimization of DNNs.

Herein, we propose a deep learning driven model termed DNNsolver, for efficient and accurate simulation of DNNs with arbitrary size and layer numbers under incident fields with arbitrary wavefronts and polarization states. Instead of learning a source-dependent mapping from structural geometry to the output field, DNNsolver learns the scattering matrix of structure, and could therefore be applied to arbitrary optical excitations. By learning local scattering within a small region and combining a sliding-window strategy with layer cascading, DNNsolver efficiently constructs the scattering response of arbitrarily sized multilayer DNNs while greatly reducing the burden of dataset generation and training. For single layer DNN, DNNsolver provides the mean square error of 0.0095 compared with FDTD for single layer DNN, while providing an approximately $10^6$-fold speedup. We further validate the model through a three-layer polarization multiplexed image classifier, with the predicted electromagnetic fields well agree with FDTD simulations.

## 2. Results

### 2.1 Principle of DNNsolver

DNNs considered in this work are composed of multiple dielectric nanopillar layers. Each nanopillar has a rectangular cross-section, giving rise to birefringence and distinct responses to horizontally (H) and vertically (V) polarized light. DNNsolver reformulates DNN simulation as the construction and cascading of source-independent local scattering responses, enabling complex multilayer optical systems to be modeled from reusable structural representations. The architecture of DNNsolver is presented in **Figure 1(a)**. We took one-dimention situation as an example for ease of illustration. The entire DNN, as a linear system, can be described by a scattering matrix $\mathbf{S} = \mathbf{F}_{\text{out}}(\mathbf{S}_M \prod_{m=1}^{m=M-1}(\mathbf{F}_{\text{m}}\mathbf{S}_m))\mathbf{F}_{\text{in}}$, where $\mathbf{F}$ denotes the free-space propagation operator and $\mathbf{S}_m$ refers to the scattering matrix of layer $m$. Although the overall scattering matrix $\mathbf{S}$ exhibits highly nonlocal coupling owing to repeated diffraction and propagation [36, 37], the scattering matrix of an individual layer is inherently sparse, with significant elements concentrated around the diagonal. This locality arises because the finite thickness of a single layer limits electromagnetic coupling to a small neighborhood. Assuming that the optical modes incident to unit $n$ of layer $m$ mainly couples to unit $n-1$, $n$ and $n+1$. the coresponding scattering coefficients $\left[S_{n,n-1}^m, S_{n,n}^m, S_{n,n+1}^m\right]$ are primarily determined by the local geometries $[P_{n-1}^m, P_n^m, P_{n+1}^m]$, where $S_{n,n'}^m$ denotes the elements in $\mathbf{S}_m$ while $P_n^m$ refers to the geometric parameters of unit $n$ in layer $m$. Based on such local correspondence, a multilayer perceptron (MLP), denoted by $\Psi_\theta$, is employed to predict the local scattering kernel from the designed geometry, where $\theta$ refers to the trainable parameters. By applying the model in a sliding-window manner across the entire layer, the complete scattering matrix $\mathbf{S}_m$ could be reconstructed. By further cascading the layers with angular spectrum method (ASM, see Methods), scattering matrix $\boldsymbol{S}$ of the entire DNN with arbitrary size is obtained.

The prediction capability of DNNsolver is illustrated in **Figure 1(b)** by a simple example. A trilayer DNN was generated randomly, with each layer containing a 30×30 array of IP-DIP nanopillars and a interlayer distance of 20 μm. The detection plane is located 20 μm behind the last layer. The output of the trilayer DNN under a H-polarized random speckle illumination at a wavelength of 1550 nm is predicted by DNNsolver. The predicted real part of electric field shows excellent agreement with the results obtained from FDTD simulation. Meanwhile, the prediction only takes about 4 ms, achieving about 450,000-fold speedup compared with 30 min taken by FDTD simulations.

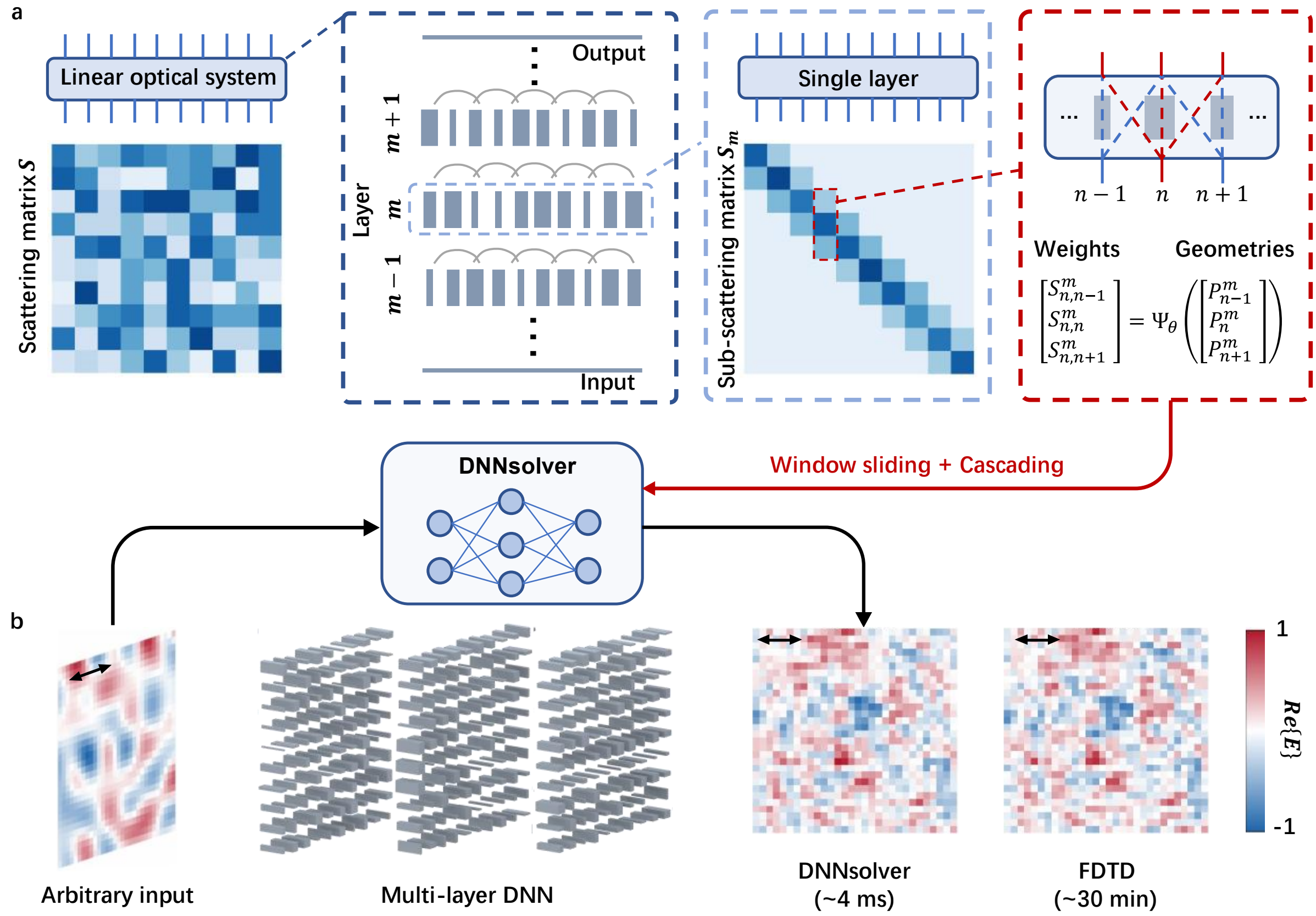

**Figure 1.** Principle of DNNsolver. a) Architecture of DNNsolver. The entire DNN is discomposed into layers with local scattering response, that learned by a lightweight MLP $\Psi_\theta$. During inference, the scattering response of an entire layer is reconstructed using a sliding-window strategy, and multilayer propagation is modeled by cascading successive layers through the ASM. b) Schematic demonstration of DNNsolver for DNN simulation. Output fields of a randomly generated trilayer DNN under random speckle illumination predicted by DNNsolver shows excellent agreement with the full-wave FDTD result.

**Figure 2** illustrates the local scattering kernel model developed in this work. The nanopillar arrays has a period of 1 μm. The height of the pillar is 4 μm, while the length $L$ and width $W$ of its cross-section ranges from 0 nm to 900 nm. A kernel size of 7×7 is adopted to define the local coupling region centered on each unit cell. The geometry of kernel is therefore described by 98 (7×7×2) parameters. After normalization to the range of [-1, 1], these parameters are flattened and fed into the MLP to predict the real and imaginary parts of the scattering weights relative to the center unit. ReLU is used as the activation function for for all hidden layers, while the output layer remains linear. It is noted that the nanopillar geometry considered in this work does not induce polarization conversion [38, 39], i.e., H- and V-polarized waves remain in their respective polarization channels after transmission.

Consequently, only the H-to-H scattering model needs to be trained, since the V-to-V response can be obtained from H-to-H response through coordinate transformation. In such case, the unit-cell array is transposed and the values of $L$ and $W$ are exchanged before being fed into the same network. The predicted output is then transposed again to obtain the corresponding V-to-V scattering response. Such symmetry allows a single network to model both polarization channels, reducing the training cost by approximately 50% while preserving the physical consistency of the predicted responses.

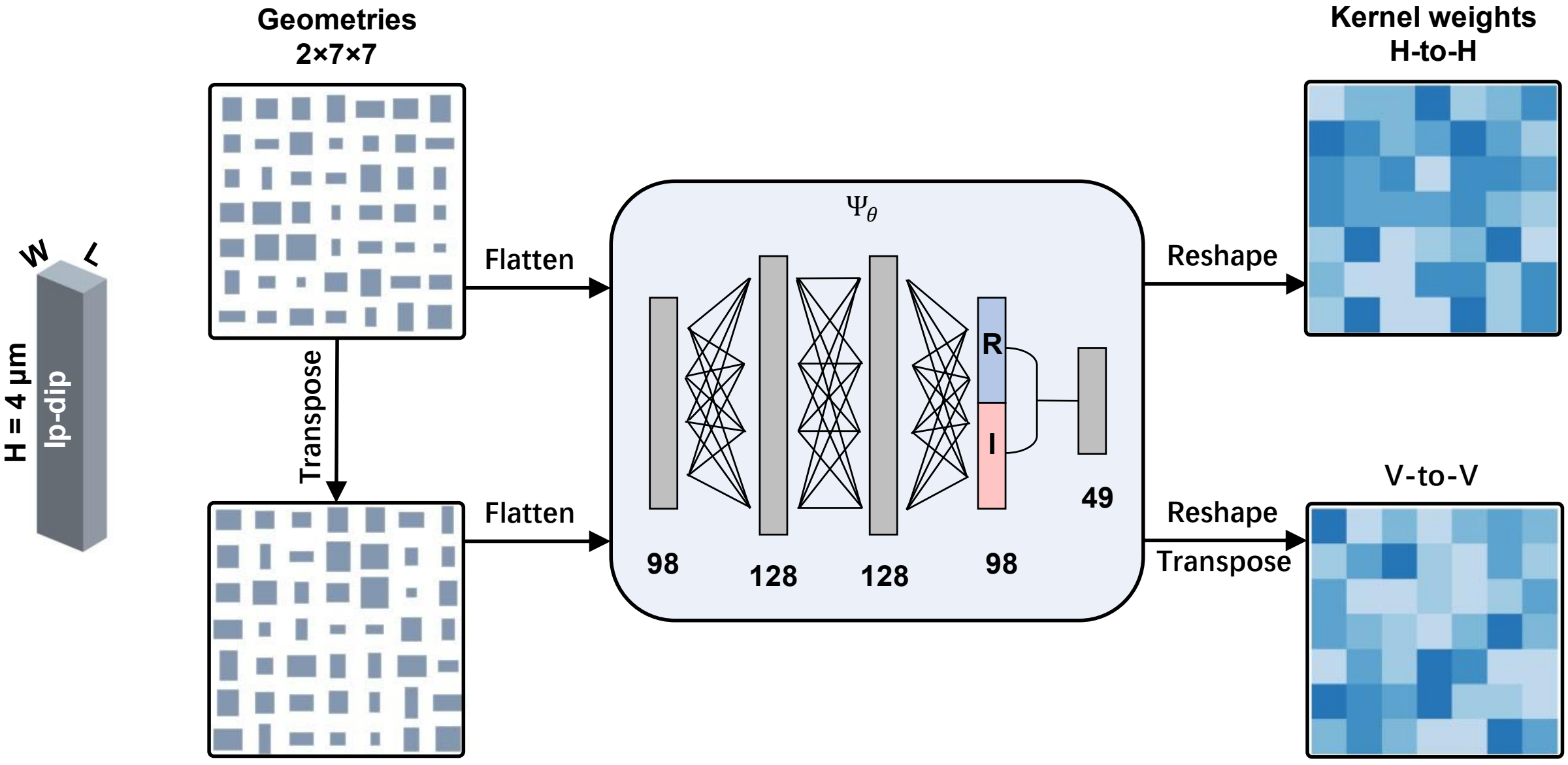

**Figure 2.** Local scattering kernel model. The MLP network $\Psi_\theta$ learns the mapping from geometries of a 7×7 local kernel to the complex-valued scattering weights associated with the central unit. With the nanopillar height fixed at 4 μm, the lengths and widths of the nanopillars constitute 98 (7×7×2) geometric input parameters. Only the H-to-H response is learned while the V-to-V response can be obtained through a coordinate-axis mirroring.

## 2.2. Training strategies and results

**Figure 3(a)** illustrates the training procedure of DNNsolver. During training, each diffractive layer consists of a 32×32 array, yielding 26×26 local kernels for a kernel size of 7×7. For dataset generation, 437 pairs of random geometries and H-polarized speckles were generated, and the corresponding near-field responses were obtained from FDTD simulations as the ground truth. In order to improve the generalizability of the model, speckles with different highest spatial-frequency components were included. All of the input speckles were normalized by the maximum amplitude. The dataset was randomly divided into training, validation, and test sets in a ratio of 0.6 : 0.2 : 0.2. The geometries were fed into the network for scattering matrix prediction, and the output field was subsequently calculated for each input speckle. The mean square error (MSE) between the model outputs and ground truth is chosen as the loss function to guide parameter optimization. The training was conducted using Python (v3.12.11) with PyTorch (v2.11.0) on a graphical processing unit (GeForce GTX3060, 6GB). The Adam optimizer was applied for the parameter optimization, and a learning rate of 0.0001 was applied. Training was completed after 2000 epochs and the convergence curve was provided in **Figure 3(b)**. When the model was applied for prediction, zero padding was used to ensure that the scattering around the edge could be predicted by the model as well (see supplementary note S1). **Figure 3(c)** compares a representative prediction from the test set with the corresponding FDTD result, revealing excellent agreement. An average MSE of 0.0095 was achieved on the test set. Furthermore, the average inference time of DNNsolver on the test set is 0.7 ms, compared with approximately 15 min for FDTD simulation. This corresponds to a speedup of more than $10^6$ over FDTD, demonstrating the high computational efficiency of DNNsolver and its potential for DNN optimization over large-scale datasets.

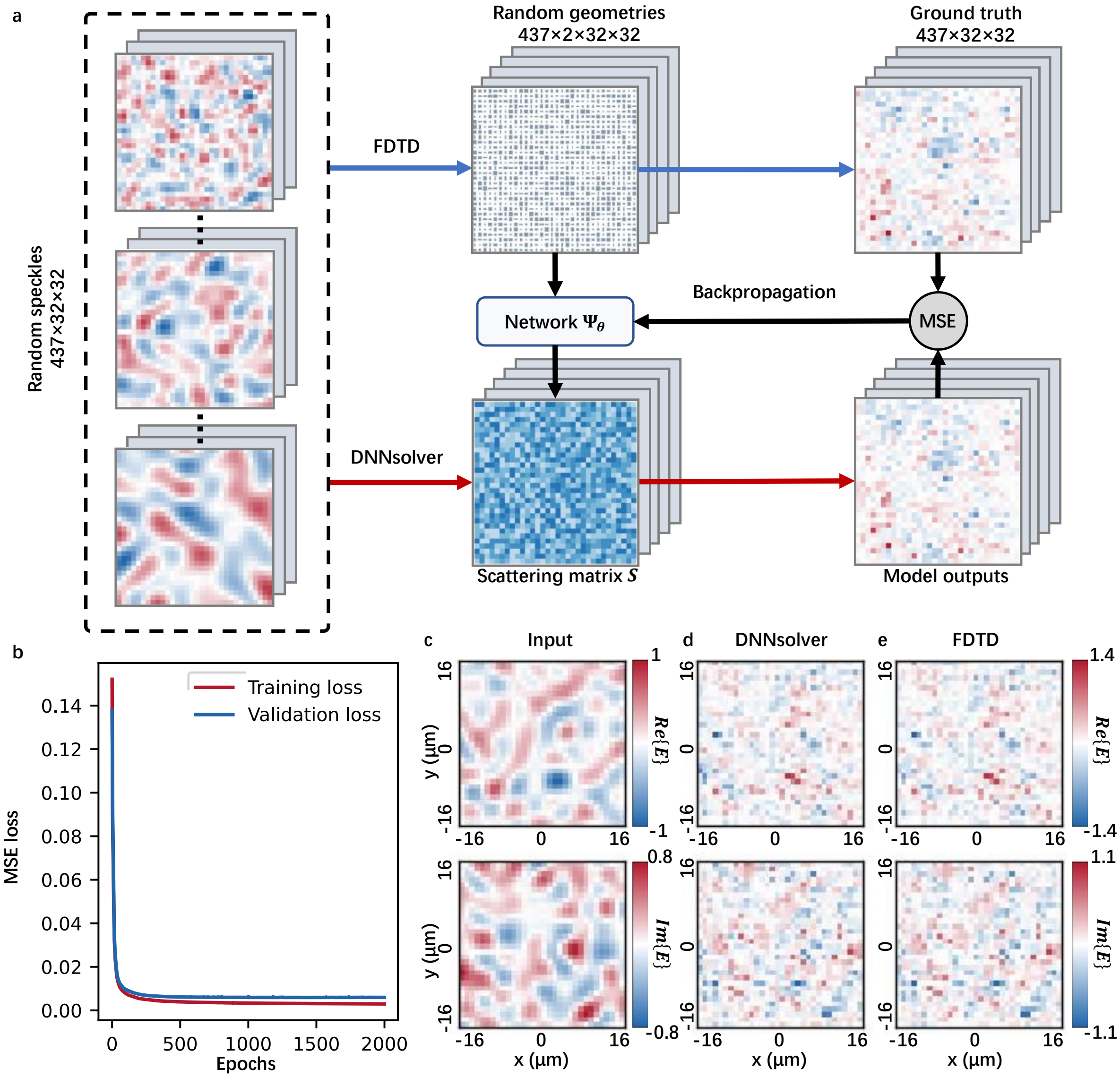

**Figure 3.** Training strategy and performance of DNNsolver. a) Dataset generation and training procedure. Random speckles with different maximum spatial frequency are included in the dataset to improve the model generalizability. The near field outputs simulated by FDTD serve as the ground truth for training. b) The MSE loss for training sets and validation sets over 2000 epochs. c-e) Real and imaginary parts of a representative input speckle (c) from the test set and the corresponding output fields predicted by DNNsolver (d) and simulated by FDTD (e). All field distributions shown are H-polarized.

## 2.3. Optimization of DNNs with DNNsolver

### *2.3.1. Polarization-multiplexed image classifier*

In order to validate the effectiveness as well as the generalizability of DNNsolver, we incorporate it into a typical design framework where DNN is optimized for a specific computation task. As a representative example, we designed a polarization-multiplexed image classifier [12]. The DNN consists of three nanopillar layers separated by 20 μm, with each layer containing a 30×30 array of nanopillars. The detection plane is located 20 μm behind the last layer. As illustrated in **Figure 4 (a)**, the DNN was designed to simultaneously classifies handwritten digits in the H-polarization channel and clothing images in the V-polarization channel. As examples, the handwritten digit '7' in H-polarization and the shoe image in V-polarization will illuminate different regions at the detection plane. The entire MNIST and Fashion-MNIST datasets were used for training. The cross entropy loss was adopted and the training stoped after 16 epoches with Adam optimizer and a learning rate of 0.1.

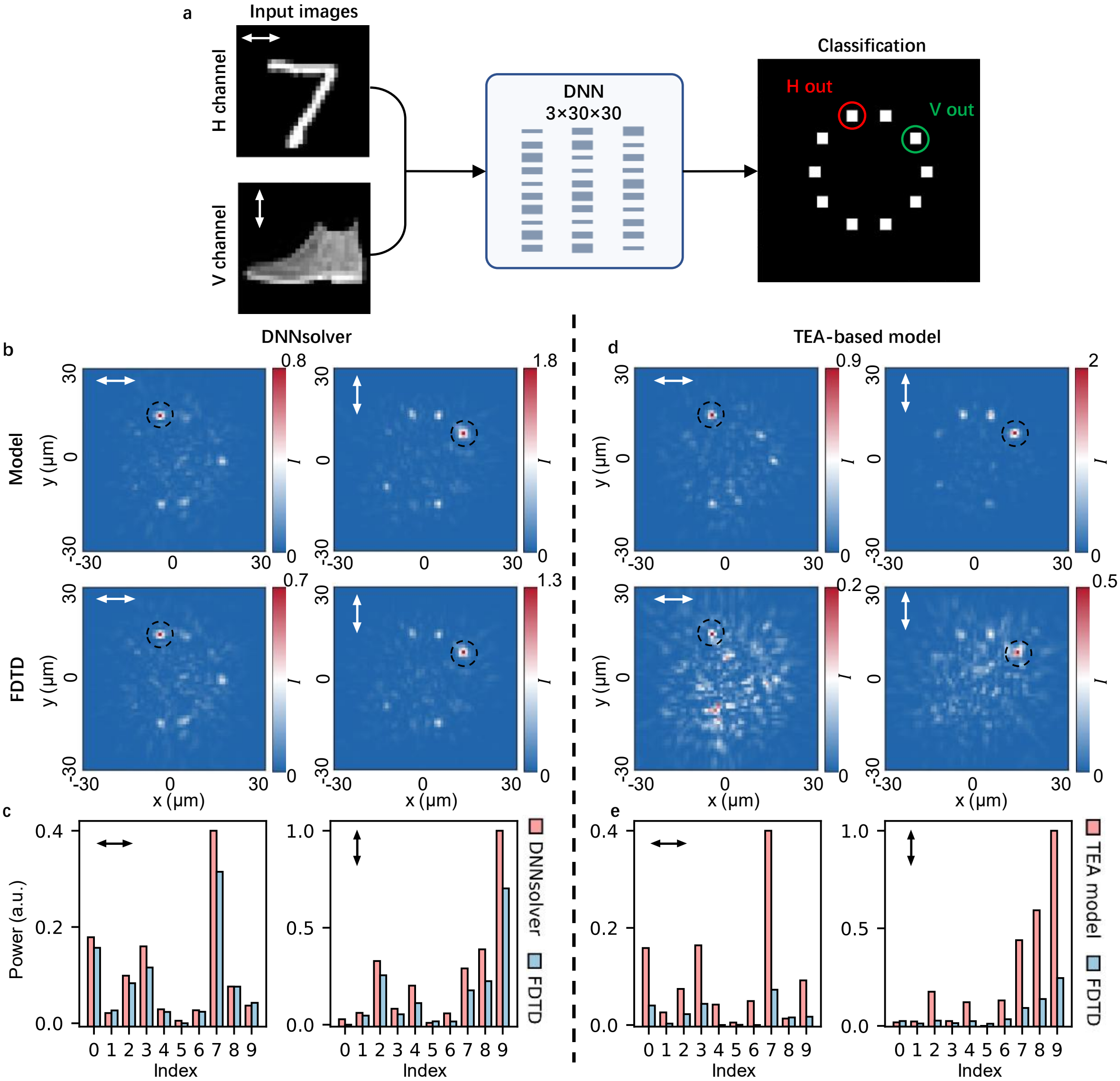

**Figure 4.** Trilayer polarization-multiplexed image classifier. a) Schematic of the device functionality. The H-polarized channel classifies handwritten digits, whereas the V-polarized channel classifies clothing images. As examples, the digit '7' in the H-channel is directed to the detection region marked by the red circle, while the shoe image in the V-channel is directed to the region marked by the green circle. b, d) Predicted and FDTD-simulated

intensity distributions for the image classifier optimized by DNNsolver (b) and TEA-based model (d). The target detection region is indicated by the black dashed circle. c, e) Optical power within ten detection regions for the image classifier optimized by DNNsolver (c) and TEA-based model (e). The power is obtained by summing the intensity within each of the ten regions marked in (a). Horizontal and vertical double-headed arrows indicate the H- and V-polarized channels, respectively.

To validate the accuracy of DNNsolver, the optimized structures were evaluated using full-wave FDTD simulations for the diffractive layers and the ASM for free-space propagation. Both the intensity distributions from the DNNsolver and FDTD are given in **Figure 4 (b)**. In addition to the target focal points (indicated by black dashed circles), the background intensity distributions are also faithfully reproduced, revealing excellent agreement between the proposed model and FDTD simulations. **Figure 4 (c)** summarizes the electric field intensity captured by ten different detection regions. Despite slight difference in part of the regions, the overall characteristics from DNNsolver remain well consistent with the results from FDTD simulations. Performance for other input images can be found in **Figure S3**. For comparison, the image classifier was also optimized using conventional model based on TEA (see supplementary note S2). As shown in **Figure 4 (d)**, although the target focal points remain correctly formed, pronounced background speckles appear in the corresponding FDTD simulations. As a result, a huge attenuation in the captured power between the TEA model and FDTD simulations was observed within ten detection regions, as shown in **Figure 4(e)**.

## 3. Conclusion

The DNNsolver proposed in this work enables efficient simulation of multilayer DNNs, with accuracy comparable to full-wave FDTD simulations. By learning the scattering matrix rather than the output field, DNNsolver decouples the incident wavefront from the model, allowing arbitrary input fields without additional training. The framework is not intrinsically constrained by the spatial size or number of diffractive layers because it learns the local scattering response instead of the entire structure. The effectiveness of DNNsolver is verified by the design of a polarization-multiplexed image classifier, for which the predicted output fields show excellent agreement with FDTD simulations in both polarization channels, whereas the DNNs designed using conventional TFA model exhibits pronounced deviations.

Although the DNNsolver is tailored for the metasurface-like layers, the methodology can be extended to any kind of layers as long as they have similar local geometries [3, 18, 34]. By introducing the rotation angle as an additional degree of freedom and expanding the training dataset, the DNNsolver framework could be further extended to polarization-sensitive DNNs with more general polarization responses. Other physical degrees of freedom, such as wavelength-dependent responses, can likewise be incorporated into the model to enable broadband optimization. DNNsolver fills the empty of a fast and accurate model for the polarization sensitive DNNs with low revractive index and high density, paving the way for high-precision design of highly integrated and multifunctional photonic devices.

## 4. Methods

*Angular spectrum method*: The free-space propagation in DNN is carried out by the angular spectrum method. For an incident vector electric field of $\boldsymbol{E}(x, y, z_0)$ and a propagation length of $d$, the output field is expressed by:

$$\boldsymbol{E}(x, y, z_0 + d) = \boldsymbol{\mathcal{F}}^{-1}\left\{ e^{i2\pi d\sqrt{\frac{1}{\lambda^2} - f_x^2 - f_y^2}} \mathcal{F}\{\boldsymbol{E}(x, y, z_0)\} \right\}$$

where $\mathcal{F}(\boldsymbol{\mathcal{F}}^{-1})$ denotes the (inverse) Fourier transform, $\lambda$ denotes the working wavelength, and $f_{x(y)}$ denotes the spatial frequency.

## Data Availability Statement

The data that support the findings of this study are available from the corresponding author upon reasonable request.

This work introduces DNNsolver, a deep-learning model that rapidly and accurately predicts light propagation through multilayer diffractive neural networks while capturing local electromagnetic coupling beyond conventional thin-element approximations. The model supports arbitrary input wavefronts and polarizations and can be applied to devices with different sizes, paving way for the accurate and efficient design of high-density, polarization-sensitive diffractive neural networks.

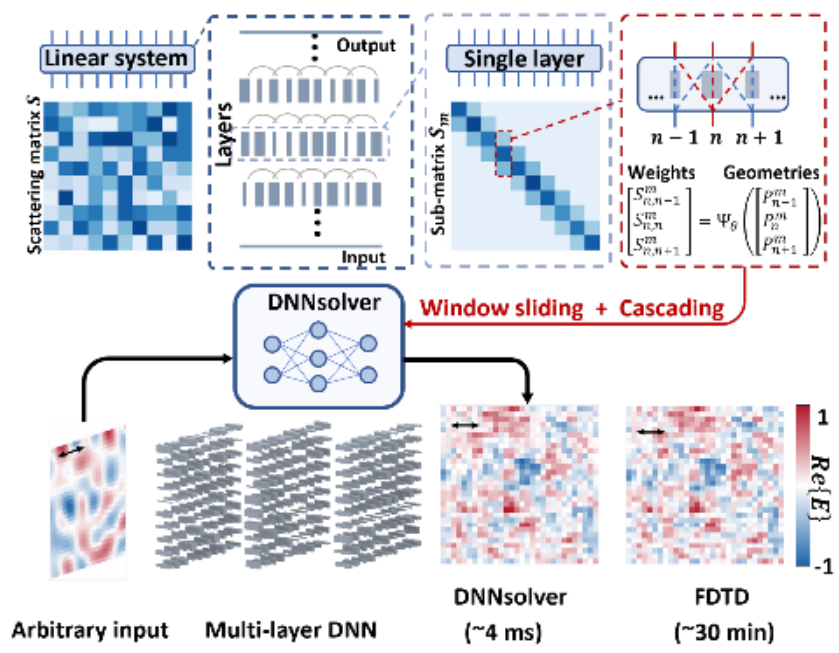